\documentclass[10pt, aps, twocolumn, nofootinbib]{revtex4-1}
\usepackage{amssymb}
\usepackage{bm}
\usepackage{graphicx}
\usepackage{array}
\newcolumntype{C}[1]{>{\centering\let\newline\\\arraybackslash\hspace{0pt}}m{#1}}

\usepackage{siunitx}
\begin{document}

\title{Density Distribution of a\\Bose-Einstein Condensate of Photons in a Dye-Filled Microcavity}

\author{S. Greveling}
\thanks{These authors contributed equally to this work}
\affiliation{Debye Institute for Nanomaterials Science $\&$ Center for Extreme Matter and Emergent Phenomena, Utrecht University, Princetonplein 5, 3584 CC Utrecht, The Netherlands}

\author{K. L. Perrier}
\thanks{These authors contributed equally to this work}
\affiliation{Debye Institute for Nanomaterials Science $\&$ Center for Extreme Matter and Emergent Phenomena, Utrecht University, Princetonplein 5, 3584 CC Utrecht, The Netherlands}

\author{D. van Oosten}
\email[Corresponding author: ]{D.vanOosten@uu.nl}
\affiliation{Debye Institute for Nanomaterials Science $\&$ Center for Extreme Matter and Emergent Phenomena, Utrecht University, Princetonplein 5, 3584 CC Utrecht, The Netherlands}

\begin{abstract}
The achievement of Bose-Einstein condensation of photons (phBEC) in a dye-filled microcavity has led to a renewed interest in the density distribution of the ideal Bose gas in a two-dimensional harmonic oscillator. We present measurements of the radial profile of photons inside the microcavity below and above the critical point for phBEC with a good signal-to-noise ratio. We obtain a good agreement with theoretical profiles obtained using exact summation of eigenstates.
\end{abstract}

\date{\today}
\maketitle

\section{Introduction}
Since the realization of Bose-Einstein condensation (BEC) in dilute atomic gases~\cite{Anderson1995,Ketterle1995}, the density distribution of BECs and their surrounding thermal cloud have been well studied using time-of-flight absorption imaging. The in-situ density distribution of a BEC in a harmonic trap is first observed using phase contrast imaging~\cite{Hulet1997}. In later work, the density distribution of BECs in a harmonic potential is also studied and has been well described theoretically using a local density approximation~\cite{Stringari1999}. This work has recently become even more relevant with the creation of exciton-polariton condensates~\cite{Kasprzak2006, Snoke2007}, which truly have a two-dimensional nature.

However, for ideal bosons the local density approximation cannot be used as the homogeneous system in that case does not show BEC at non-zero temperature. This problem has recently become  relevant with the achievement of phBEC~\cite{Klaers2010}, which are expected to behave as ideal bosons. Initial theoretical work on the density distribution of phBECs~\cite{Kruchkov2014}, which relies on the local density approximation, should therefore be used with caution.

In this work we study the density distribution of photons in a dye-filled microcavity, both experimentally and theoretically. Experimentally, we exploit the axial symmetry of the system by computing radial averages of the photon gas yielding a good signal-to-noise ratio of the overall signal, especially in the thermal tail of the distribution. This allows us to accurately study the tail and compare the behavior of the tail below and above the phase transition. Theoretically, we calculate the radial density profiles by carrying out an exact summation of harmonic oscillator states weighed by the appropriate Bose-Einstein distribution function. We find good correspondence between theory and experiment. We also find, both experimentally and theoretically, that below the phase transition the thermal distribution has the expected Gaussian form, but that above the transition the thermal tail becomes strongly non-Gaussian.
\section{Experimental setup}  \label{sec:Chap4_exp_setup}
The optical design of our experimental setup is based on the work of the Klaers~\textit{et al.} \cite{Klaers2010, Klaers2011}. The core of our experimental setup is a cavity consisting of two ultra-high reflecting, spherical mirrors with a radius of curvature of \SI{1}{\meter} with a separation on the order of \SI{1}{\micro \meter}, between which a droplet of a Rhodamine 6G dye solution is placed. The dye solution is pumped by laser pulses with a duration of \SI{500}{\nano \second} and a wavelength of 532 nm, which are created using a CW laser and acousto-optic modulators (AOMs). Using three AOMs in series gives us an extinction ratio of \num{5E4}, ensuring a high on/off contrast of the pump pulse. 

The density distribution of photons escaping through one of the cavity mirrors is imaged. The photons first pass through a lens with a diameter of \SI{25.4}{\milli \meter} placed at the focal distance $f_{1} = \SI{75}{\milli \meter}$ away from the cavity. The consequences of the effective aperture of this collecting lens will be discussed later. After passing through the lens, a second lens with a focal length $f_{2} = \SI{200}{\milli \meter}$ is used to created an image plane with a magnification $M_{1}= f_{2}/f_{1} = \num{2.67}$.
\begin{figure}[!b]
  \centering
  \includegraphics[width=0.95\linewidth]{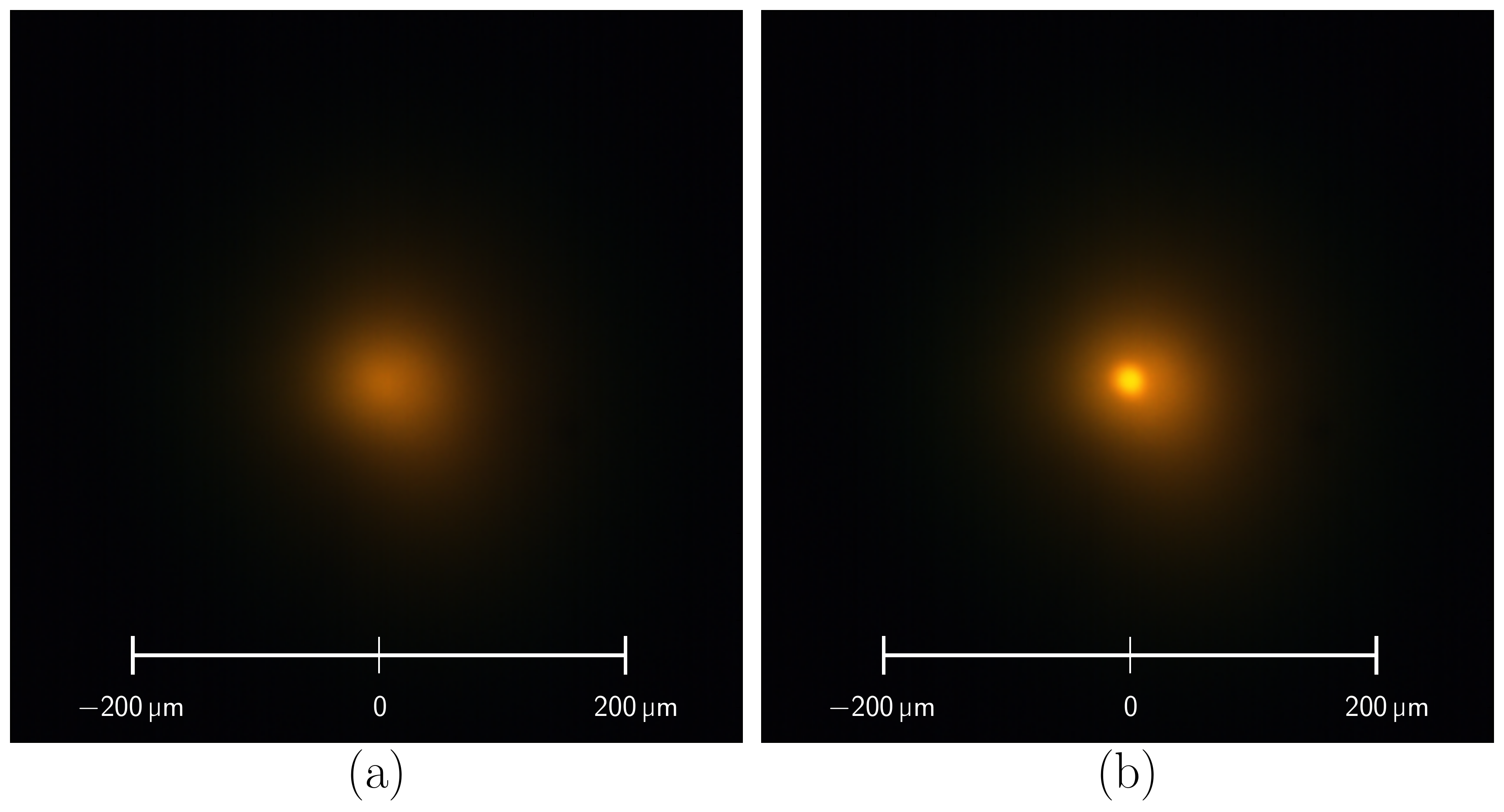}
  \caption{Typical images of the photon distribution inside the cavity below (a) and above (b) the critical point for phBEC. The bright core in the right image is the Bose-Einstein condensate, whereas the surrounding distribution is the thermal cloud.}
\label{fig:typical}
\end{figure}

Figure~\ref{fig:typical} shows typical images obtained below and above the BEC phase transition acquired using an RGB color camera in the image plane. During the experiments the RGB camera is removed and the image plane is imaged on a high resolution camera using a second set of lenses, yielding an additional magnification $M_{2} = \num{6}$. The camera and the AOMs are synchronized such that we can image the distribution of photons in the cavity during each individual pump pulse. To ensure reproducibility of our results, we perform our experiment in runs during which thousands of images are taken without user intervention. 
\subsection{Automation}
The heart of the automation system of the experiment is the Beaglebone Black (BBB), a low-cost development platform built around the ARM Cortex-A8 processor (Sitara), running a Linux operating system. The BBB is particularly suitable for real-time applications, because the Sitara processor includes two so-called programmable real-time units (PRUs), which share memory with the ARM core. We use one such PRU to address a serial ten bit digital-to-analog converter (DAC). The voltage from the DAC is used to control the RF power of one of the AOMs, allowing us to change the pump pulse power during the experimental run. Furthermore, the PRU is programmed to provide trigger pulses for the AOMs and the camera. The PRU is instructed to trigger a shot of the experiment by sending the appropriate commands over shared memory using a Python script running on the ARM core.

The camera is an Andor Zyla 5.5 sCMOS, the data of which are read out by a PC over a USB3 bus. At the start of an experimental run the camera is set up for external triggering and a Python script runs that takes the appropriate number of images. Afterwards, we start a Python script on the BBB that sends the corresponding shot-requests to the PRU. Synchronization is maintained by the fact that the camera is externally triggered by the PRU. 
 
An experimental run typically consist of 50 repetitions of the same sequence containing typically \num{60} shots each with a different pump power. Additionally, for each shot a background image is taken. During a run we take 8 shots per second, which including the background images leads to a frame rate of 16 frames per second (fps). As a single full frame image of the camera contains approximately $11\,\mathrm{MBytes}$, this corresponds to a data rate of $88\,\mathrm{MBytes}/\mathrm{s}$. The total amount of data produced in one run is $66\,\mathrm{GBytes}$. To handle the data flow, we temporarily store the data of \num{10} experimental runs on a solid-state-drive (SSD), with a capacity of $1\,\mathrm{Tbytes}$. At the end of the day we copy the data from the SSD to other storage media during the following night.

As the experiment runs completely without user intervention during an experimental run, the lab is vacated during runs, which provides additional stability and reproducibility contributing to the quality of our data.
\section{Theory}
In the grand canonical ensemble, the total number of identical, ideal bosons in a trap with a set of discrete energy levels $E_{n}$ is given by
\begin{equation}    \label{eqn:particle_number}
  N = \sum_{n} g_{n} f_{\mathrm{BE}}(E_{n}, \mu, k_{\mathrm{B}} T),
\end{equation}
where $f_{\mathrm{BE}}$ denotes the well-known Bose-Einstein distribution function, $g_{n}$ the degeneracy of the $n$-th energy level, $\mu$ the chemical potential, $k_{\mathrm{B}}$ the Boltzmann constant, and $T$ the temperature. In our dye-filled microcavity, the energy levels are those of the isotropic two-dimensional quantum harmonic oscillator. The principal quantum number is $n = n_{x} + n_{y}$, with $n_{x, y} \in \mathbb{N}_{0}$ and $E_{n_{x} n_{y}} = \hbar \Omega (n_{x} + n_{y} + 1)$, where $\Omega$ denotes the harmonic oscillator frequency. We assume that neither the cavity nor the medium inside the cavity exhibits birefringence,~\textit{i.e.} that the harmonic oscillator frequency does not depend on the polarization. Summation over the polarization thus yields a factor \num{2}.

To determine the density of photons inside the cavity, we substitute the probability distribution of each state of $\left(n_{x}, n_{y}\right)$ into the above summation,~\textit{i.e.}
\begin{equation}
  \rho(x, y) = 2 \sum_{n_{x}, n_{y}} \left| \psi_{n_{x}}(x) \psi_{n_{y}}(y)\right|^{2} f_{\mathrm{BE}}(E_{n_{x} n_{y}}, \mu, k_{\mathrm{B}} T),
\end{equation}
where the functions $\psi_{n}(x)$ are the well-known one-dimensional harmonic oscillator wave functions. Because the expression contains a double summation over a potentially large number of states, it is unfeasible to use in a least-square fitting procedure. For that reason, we write the density as
\begin{equation}
  \rho(x, y) = \sum_{n} g_{n} \rho_{n}(x,y) f_{\mathrm{BE}}(E_{n}, \mu, k_{\mathrm{B}} T),
\end{equation} 
where $g_{n} = 2(n + 1)$ denotes the degeneracy and,
\begin{equation}
  \rho_{n}(x, y) = \frac{1}{n + 1} \sum_{n_{x} = 0}^{n} \left| \psi_{n_{x}}(x) \psi_{n - n_{x}}(y)\right|^{2}.
\end{equation}

As the above expression for $\rho_{n}(x, y)$ does not depend on the parameters $\mu$ and $T$, we can precompute the summation over $n_{x}$ and carry out the summation over $n$ during the fitting procedure. An added benefit of this method is that in the isotropic system at hand, $\rho_n$ can only depend on the radial coordinate $r$,~\textit{i.e.} it must be radially symmetric. Thus we can compute $\rho_{n}(r)$ by computing the angular average of $\rho_{n}(x, y)$ and use
\begin{equation}    \label{eq:Chap4_radprof_theory}
  \rho(r) = \sum_{n} g_{n} \rho_{n}(r) f_{\mathrm{BE}}(E_{n}, \mu, k_{\mathrm{B}} T) \eta(E_{n}),
\end{equation} 
where we have introduced the detection efficiency $\eta(E_{n})$ that depends on the photon energy, due to for instance the quantum efficiency of the camera used in the experiment.
\subsection{Implementation}
Even though the eigenfunctions of the one-dimensional quantum harmonic oscillator are analytically known, their computation presents a numerical problem. For quantum numbers from $n\approx 150$ and upwards the computation involves intermediate values that exceed the maximum value of the double format. For this reason, we calculate the harmonic oscillator wave functions using the Numerov method~\cite{Johnson1977}. In order to reach a sufficient accuracy for the large quantum numbers, we use a grid with a resolution of $\Delta x = \num{2E-3} \, l_{\mathrm{HO}}$, where $l_{\mathrm{HO}}$ denotes the harmonic oscillator length. We are left with an array $\psi_{nm}$, where $n$ is the quantum number and $m$ labels the position according to $x_{m} = -50 l_{\mathrm{HO}} + m \Delta x$. We subsequently take the absolute value squared of the numbers in this array, apply a low-pass filter along the $m$-direction and sub-sample the array in the $m$-direction to account for the finite optical resolution. Using the resulting one-dimensional density distributions, we tabulate the functions $\rho_n(r_m)$, where $r_m=m \Delta x$. To use the tabulated data in the fit procedure, we create interpolating functions that allow us to scale the harmonic oscillator length to the experimental value.
\begin{figure}[!b]
  \centering
  \includegraphics[width=0.95\linewidth]{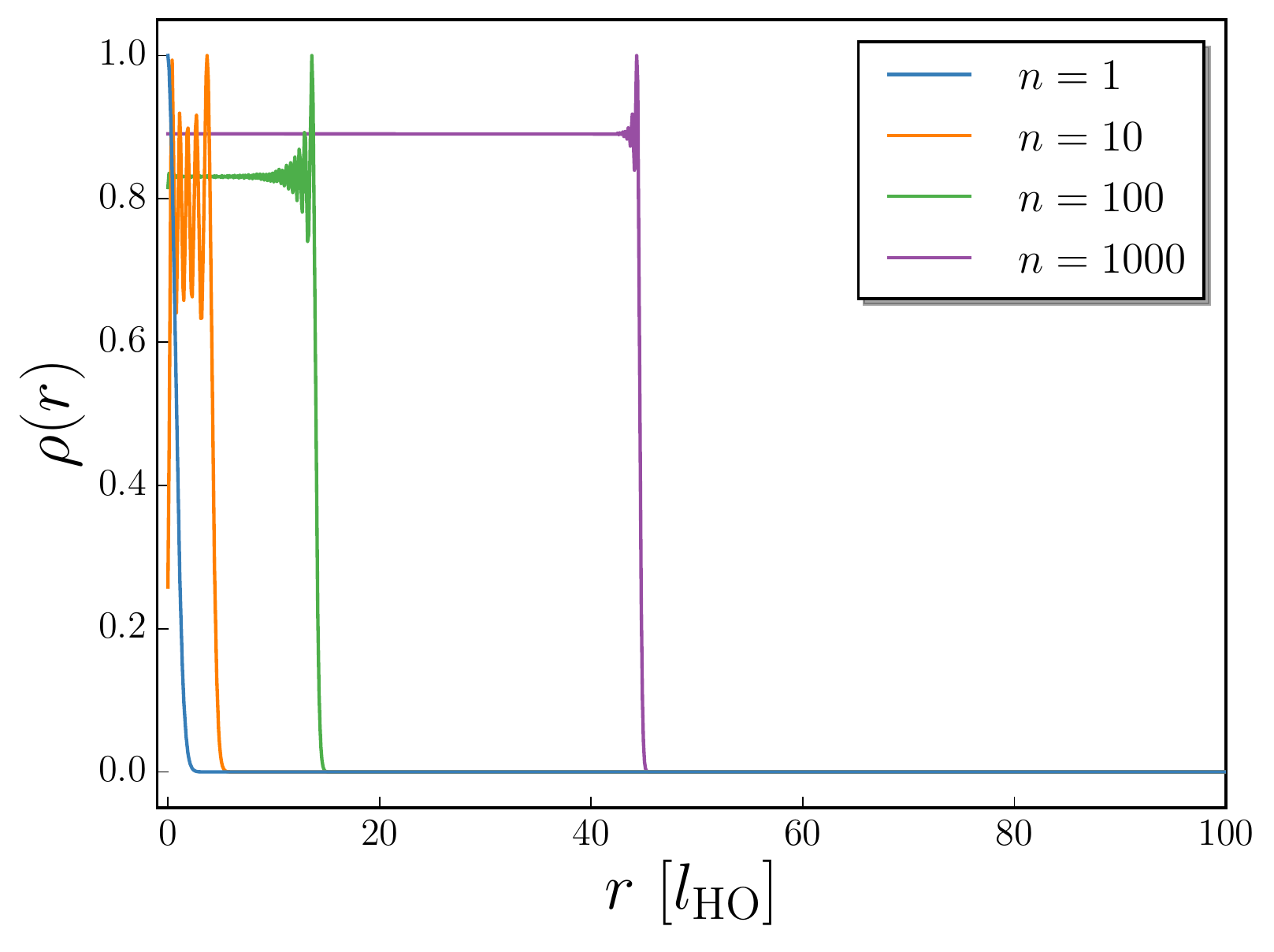}
  \caption{Typical examples of $\rho_{n}(r)$ normalized to the maximum value.}
\label{fig:rhons}
\end{figure}

Examples of the resulting functions are shown in Fig.~\ref{fig:rhons}. For $n = 1$ we find the familiar Gaussian form. For larger principle quantum numbers the density distribution rapidly oscillates, as expected. Due to the filtering, the fast oscillations are suppressed. For principle quantum numbers in the order of \num{100} or \num{1000}, the filtered radial distribution behaves more and more like a step function, with only a fast oscillation around the classical turning point $r_{\mathrm{ctp}}= \sqrt{2n} \, l_{\mathrm{HO}}$. Therefore, for principle quantum numbers $> \num{2000}$, we no longer perform the numerical calculation, but rather use an appropriately normalized step function.

\section{Results}       \label{sec:Chap4_Results}
\subsection{Experimental profiles}
To quantitatively analyse the results, we exploit the fact that the photon distributions are radially symmetric, as the microcavity is isotropic, allowing us to perform a radial average. We first determine the center of the microcavity by selecting a run with a visible condensate and locate the center pixel of the condensate. Performing the procedure for several randomly selected images, we notice that we always find the same center pixel, showing the stability of the setup. From the center pixel we average each image radially outwards, allowing us to increase the overall signal-to-noise ratio, but especially in the tail of the distribution that is otherwise very noise.
\begin{figure}[!b]
  \centering
  \includegraphics[width=0.95\linewidth]{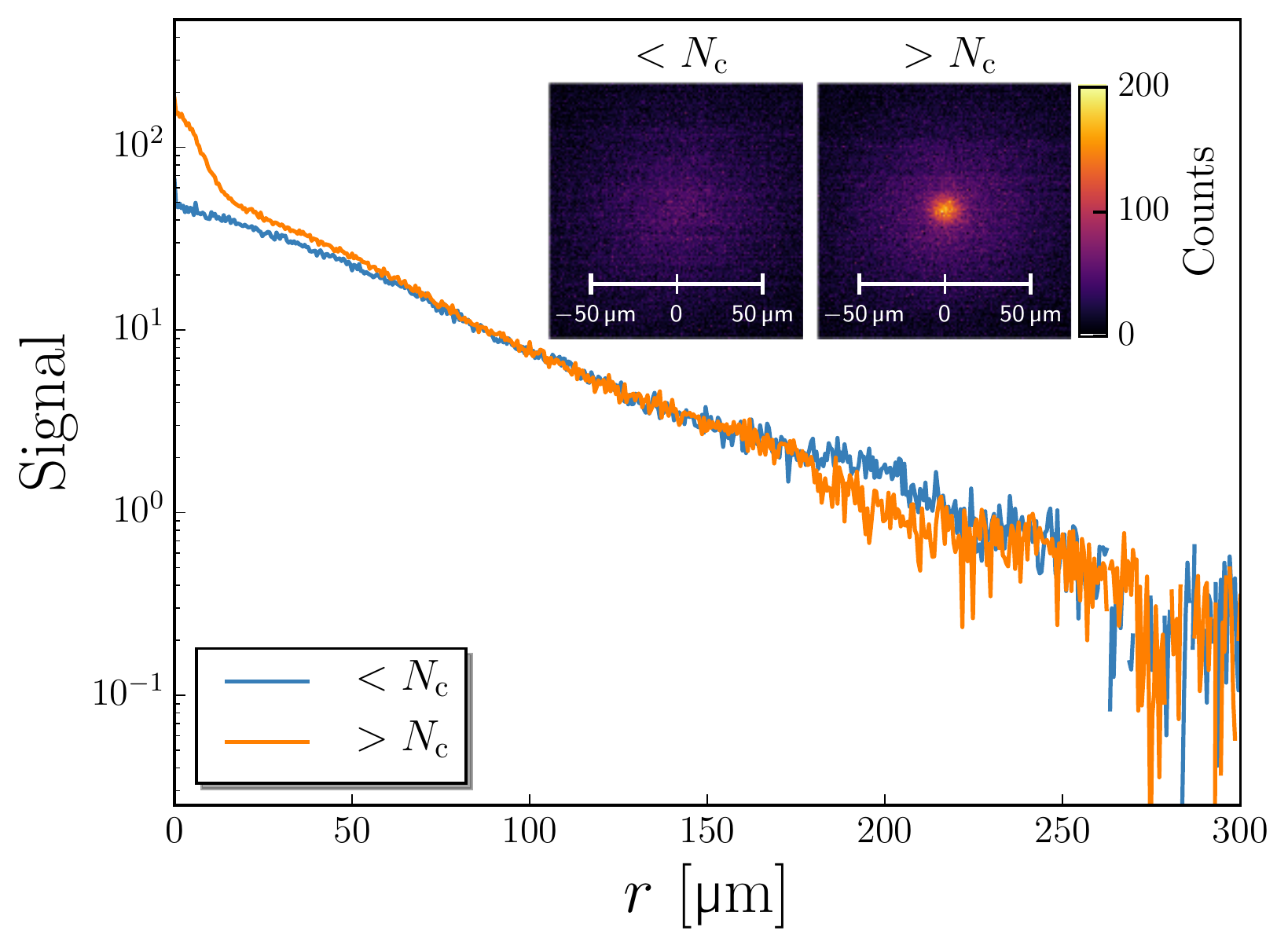}
  \caption{Examples of experimental radial profiles measured below (blue) and above (orange) the critical point for phBEC. We can detect the decrease of intensity in the thermal tail over more than two orders of magnitude. The insets show the (false color) camera images, from which these radial profiles are computed.}
\label{fig:exp_typical}
\end{figure}

In Fig.~\ref{fig:exp_typical} we plot examples of experimentally obtained radial profiles, below and above the critical photon number $N_{c}$. The two profiles overlap except close to the center, where the orange line has an additional peak. The additional peak is caused by the macroscopic occupation of the ground state,~\textit{i.e.} the phBEC. The overlap between the profiles for larger distances, is a direct consequence of Bose-enhancement; the thermal cloud is saturated and thus cannot contain more photons. Additional photons must occupy the ground state.

One can also observe that the profiles deviate from each other close to, but outside the condensate. This is a consequence of the fact that we are dealing with a finite system and that we are thus not formally in the thermodynamic limit. 

\begin{figure}[!b]
  \centering
  \includegraphics[width=0.95\linewidth]{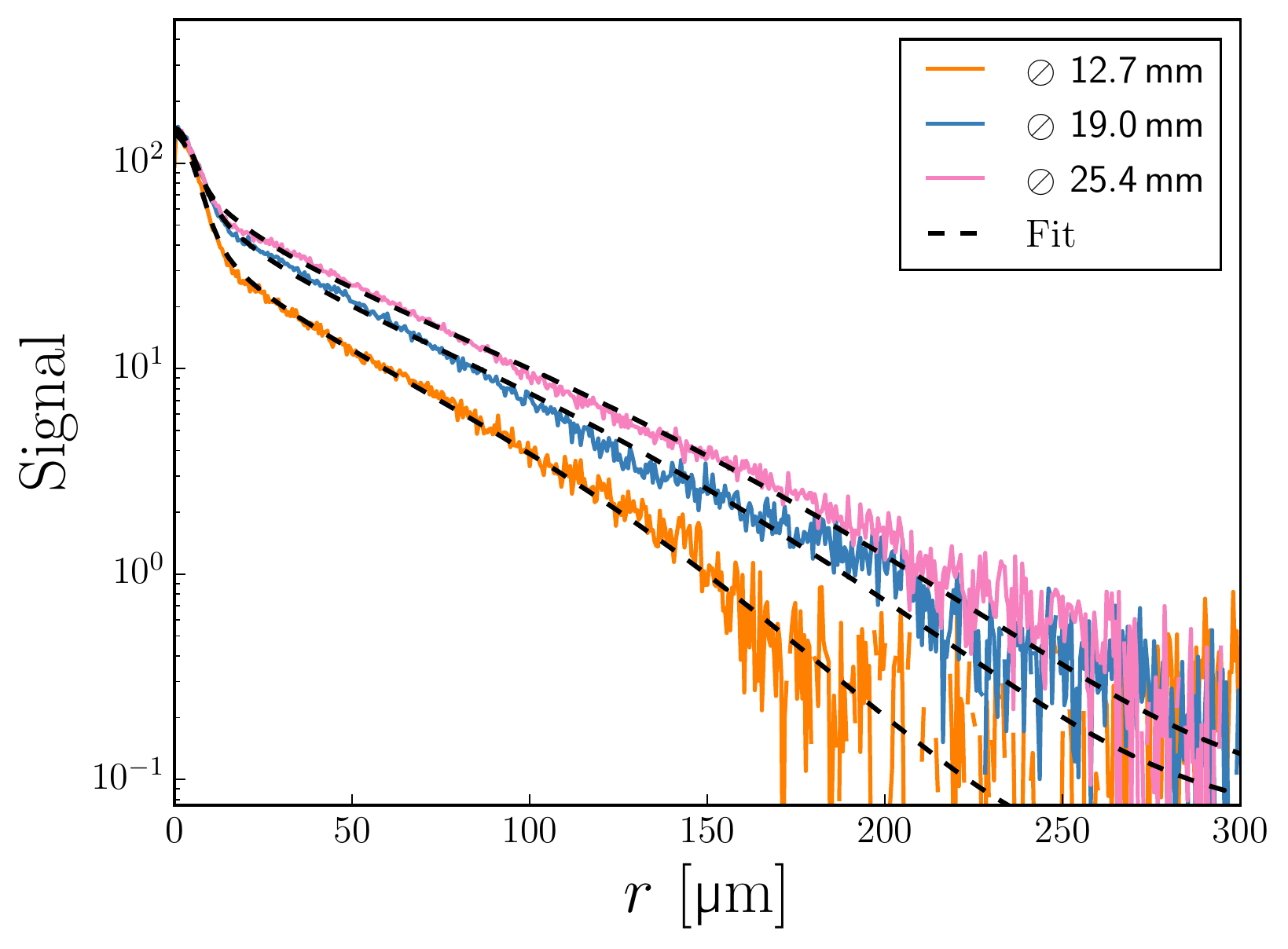}
  \caption{Radial averages for measurements under identical pumping conditions, but with a varying numerical aperture for the collecting lens. We see that reducing the numerical aperture clearly cuts the thermal tail. The dashed lines show fits to our theoretical model.}
\label{fig:exp_aperture}
\end{figure}

Lastly, we observe the decrease of the thermal tail of the distribution over more than two orders of magnitude. It is crucial to use a collecting lens with a sufficiently large numerical aperture, as illustrated in Fig.~\ref{fig:exp_aperture}. The numerical aperture of the collecting lens is reduced by inserting an aperture in front of the lens. Reducing the aperture drastically reduces the thermal tail, which has a large impact on the analysis. First, the condensate fraction is over-estimated, as most of the thermal photons are not recorded. Second, as the slope of the thermal tail is a measure of the temperature, a smaller aperture produces a steeper slope and thus predicts a lower temperature. As seen from the figure, the difference between an aperture of \SI{19.0}{\milli \meter} and \SI{25.4}{\milli \meter} is minor, which suggests that a \SI{25.4}{\milli \meter} aperture is sufficient. 
\begin{table}[!b]
  \centering
  \begin{tabular}{| C{0.17\linewidth} | C{0.27\linewidth} | C{0.19\linewidth} | C{0.19\linewidth} | }
        \hline
      $\oslash$                 & $\mu\,\,[\hbar \Omega] $ & $T\,\,[\hbar \Omega]$ & $T\,\,[\si{\kelvin}]$ \\
        \hline
      \SI{12.7}{\milli \meter}  & \num{-0.181(3)} & \num{77.5(2)}         & \num{122.1(3)}  \\
      \SI{19.0}{\milli \meter}  & \num{-0.278(6)} & \num{106.9(2)}        & \num{168.5(4)}  \\
      \SI{25.4}{\milli \meter}  & \num{-0.42(1)} & \num{123.1(2)}         & \num{194.0(4)}  \\
        \hline
    \end{tabular} 
    \caption{Fit results of the radial profiles for different apertures.}
\label{table:fit_parameters}
  \end{table}
\subsection{Fitting}
The theoretical radial profiles depend on several parameters. First of all, we need to know the harmonic oscillator length, which one can obtain from the cutoff wavelength $\lambda_{\mathrm{cutoff}}$, using~\cite{Klaers2011}
\begin{equation}
  l_{\mathrm{HO}} = \sqrt{\frac{1}{4 \pi} \frac{\lambda_{\mathrm{cutoff}}}{n_{0}}}\sqrt[4]{q R \frac{\lambda_{\mathrm{cutoff}}}{n_{0}}},
\end{equation}
where $n_{0}$ denotes the refractive index of the solvent, $q$ the longitudinal mode number of the cavity mode and $R$ the radius of curvature of the mirrors. To relate the theoretical profile to the experimental profile, the detection efficiency needs to be determined which takes into account the quantum efficiency of the camera, and the reflection coefficients of the optical elements in the experimental setup. To determine the efficiency, we use the fact that at the phase-transition the total number of photons is given by Ref.~\cite{Klaers2011}:
\begin{equation}
  N_{c} = \frac{\pi^{2}}{3} \left(\frac{k_{\mathrm{B}T}}{\hbar \Omega}\right)^2,
\end{equation}
where $\Omega$ can be expressed in the experimental parameters as
\begin{equation}
  \Omega = \frac{c}{n_{0}} \sqrt{\frac{2}{\sqrt{q R (\lambda_{\mathrm{cutoff}}/n_{0})}}}.
\end{equation}
Using an iterative method, we find the detection efficiency by dividing the value of $N_{c}$ by the total experimental signal $N_{\mathrm{exp}}$, which we obtain by numerically integrating the experimentally obtained radial profile. However, we first need to obtain the temperature of the photon gas at the critical point. We therefore fit the radial profile using the chemical potential $\mu$ and the temperature $T$, both in units of $\hbar \Omega$, as fit parameters. For the fit, we temporarily use the detection efficiency as a free parameter. With the fitted temperature, we can then recalculate the detection efficiency. The fitted and recalculated detection efficiencies agree reasonably well. We repeat the procedure for \num{50} images at the phase-transition of Bose-Einstein condensation, and average the recalculated detection efficiencies. The average detection efficiency is used and kept fixed while we carry out the fits to the entire data set.

Figure~\ref{fig:exp_aperture} shows examples of fits as black dashed lines to the radial profiles. One can note the striking agreement between the fit and the large aperture data over many orders of magnitude in signal, both for the thermal cloud and the Bose-Einstein condensate. The fit parameters we find for the different apertures are given in Table~\ref{table:fit_parameters}. We notice that the fitted temperature, when converted to SI units, is not equal to room temperature as we would have expected. This could be due to the aperture of the collecting lens. As mentioned in Sec.~\ref{sec:Chap4_exp_setup}, we image the cloud using a lens with a finite aperture. Limiting the aperture suppresses the thermal tail, but more importantly increases the slope. One can see that this plays a role, as the temperature for the smaller apertures deviates further from room temperature.
\section{Conclusion}
In our system, a Bose-Einstein condensate of photons in a dye-filled microcavity, the signal-to-noise ratio is improved by radially averaging the data. This is especially true for the tails of the thermal distributions; a crucial step as it is the tail that gives a reliable measurement of the temperature of the system. Furthermore, we show that for the same reason, it is crucial to have a large numerical aperture in the imaging system. Finally, we present a theoretical model that we use to fit our data and demonstrate a good correspondence between this model and our data.
\section*{Acknowledgements}
It is a pleasure to thank Erik van der Wurff, Henk Stoof, and Peter van der Straten for useful discussions. This work is part of the Netherlands Organization for Scientific Research (NWO).

\end{document}